\documentclass[prd,aps,preprintnumbers,showpacs,10pt]{revtex4}
\usepackage{graphicx}
\usepackage{epstopdf}
\usepackage{color}
\usepackage{epsfig}
\usepackage{subfigure}

\begin{document}
\preprint{BARI-TH/605-09}
\title{Doubly heavy baryons in a Salpeter model  with AdS/QCD inspired potential}
\author{\textbf{Floriana Giannuzzi}}

\affiliation{Universit\`a degli Studi di Bari, I-70126 Bari, Italy\\ INFN, Sezione di Bari, I-70126 Bari, Italy}
\begin{abstract}
The spectrum of baryons with two heavy quarks is predicted, assuming a  configuration of a light quark and a heavy diquark. The masses are computed within a semirelativistic quark model, using a potential obtained in the framework of the AdS/QCD correspondence. All the parameters defining the model are determined fitting the meson spectrum. The obtained mass of $\Xi_{cc}$ is in agreement with the measurements.
\end{abstract}

 \maketitle

Doubly heavy baryons, {\it i.e.} baryons made up of two constituent heavy quarks and a light quark, are predicted to exist by the quark model \cite{Fleck:1988vm}. However, the only state observed so far is a candidate for  $\Xi_{cc}$ reported  by the SELEX Collaboration, which found a signal  for the decay $\Xi_{cc}^+\rightarrow\Lambda_cK^-\pi^+$ \cite{Mattson:2002vu}. The same Collaboration confirmed   the production of $\Xi_{cc}^+$ considering the decay mode $\Xi_{cc}^+\rightarrow pD^+K^- $  \cite{Ocherashvili:2004hi}, with measured mass of $\Xi_{cc}$:
\begin{equation}\label{csiccselex}
M_{\Xi_{cc}}=3518.9 \pm 0.9 \mbox{ MeV .}
\end{equation}
 Although at present no other experiment has observed such  hadrons, it is possible that forthcoming analyses at LHC and Tevatron \cite{Kiselev:2001fw} and the future experiments like PANDA at GSI could be able  to observe the production and  decays of doubly heavy baryons. These particles deserve attention since, as pointed out  in ref. \cite{Liu:2007twb},  the observation at LHCb of the decays of either $\Xi_{cc}$ to charmless final states or $\Xi_{bb}$ to bottomless final states  would be a signal for new Physics, being  these processes strongly suppressed in the Standard Model.

In the quark model baryon spectroscopy has been discussed following two different approaches. One investigates  the three-body problem of the bound state of three quarks. The other one is based on the hypothesis, introduced in \cite{Ida:1966ev}, that  a diquark can form inside the baryon, thus reducing the description to  a two-body problem of the bound state of a diquark and a quark (for a recent review see \cite{Klempt:2009pi}).

This paper follows the second approach, supposing  that a baryon can be treated analogously to a $\bar q q$ system made up of a diquark and a quark.  This idea comes from the observation, in group theory, that two quarks can attract one another in the $\bar{3}$  representation of $SU(3)_{color}$, thus forming a diquark having the same color features as an antiquark. This suggests that the interaction between a quark and a diquark inside a baryon can be studied in analogous way as the one between a quark and an antiquark inside a meson. 
However, this does not imply that a baryon really has this structure:  the issue is still debated and one can consider this idea as the starting point for the description of baryons. \\
In particular, the system where such an idea should be properly applied is the one we are considering here, namely the baryon where two heavy quarks form a heavy diquark acting as a static color source for the third constituent light quark.
 In fact, one expects that the two heavy quarks are very close, in such a way that they are seen as a whole system by the light quark. Heavy particles are also the best objects to deal with in the model described in this paper, since it involves a static potential. 
The model was introduced in  ref. \cite{Carlucci:2007um,Giannuzzi} to compute the spectrum of heavy mesons and it is based on a semirelativistic wave equation, the Salpeter equation, with a static potential, whose eigenvalues are the masses of the bound states. 
 
 In order to compute baryon masses in this approach, there are three steps to follow.
 
  The first step is to compute heavy diquark masses. A diquark is a bound state of two interacting quarks, and the energy of this pair is, in the  one-gluon-exchange approximation, one half of the energy of a quark-antiquark pair $V(r)$. Diquark masses can be computed solving the Salpeter equation (we consider the $\ell$=0 case):
 \begin{equation}\label{salpetereq}
\left(\sqrt{m_1^2-\nabla^2}+\sqrt{m_{2}^2-\nabla^2}+\frac{1}{2}{V}(r)\right)
\psi_d({\bf r})\,=\,M_d\, \psi_d({\bf r})\, ,
\end{equation}
 where $m_1$ and $m_2$  are the masses of the quarks,  $M_d$ and $\psi_d({\bf r})$ are the mass and the wave function of the diquark, respectively,  and
 \begin{equation}\label{potenziale}
V(r)=V_{AdS/QCD}(r)+V_{spin}(r) \,.
\end{equation}
In (\ref{potenziale}), $V_{AdS/QCD}(r)$ describes the color interaction between a  quark and an antiquark, while the factor 1/2 in (\ref{salpetereq}) accounts for the quark-quark interaction in the $\bar 3$. The expression for $V_{AdS/QCD}$, apart from a constant term $V_0$, has been obtained in a gauge/gravity  framework in ref. \cite{Andreev:2006ct} in a parametric form:
\begin{equation}\left\{
\begin{array}{cc}
\label{potadsqcd} 
V_{AdS/QCD}(\lambda)\,=\,\frac{g}{\pi}
\sqrt{\frac{c}{\lambda}} \left( -1+\int_0^1 dv \, v^{-2} \left[
\mbox{e}^{\lambda v^2/2} \left(1-v^4
\mbox{e}^{\lambda(1-v^2)}\right)^{-1/2}-1\right]\right) & \\
r(\lambda)\,=\,2\, \sqrt{\frac{\lambda}{c}} \int_0^1 dv\, v^{2}
\mbox{e}^{\lambda (1-v^2)/2} \left(1-v^4
\mbox{e}^{\lambda(1-v^2)}\right)^{-1/2}  \hspace{3.3cm}&  \,,
\end{array}
\right.\end{equation}
where $r$ is the interquark distance and $\lambda$ varies in the range: $0\leq\lambda<2$.
The term $V_{spin}(r)$  accounts for the spin interaction, and is given by: 
\begin{equation}
V_{spin}(r)\,=\,A \frac{\tilde\delta(r)}{m_1 m_2}{\bf S_1}\cdot{\bf
S_2} \qquad\quad\mbox{with }\qquad \tilde\delta(r)=\left(\frac{\sigma}{\sqrt{\pi}}\right)^3
e^{-\sigma^2 r^2}\,, 
\end{equation}
where $\sigma$ is a parameter defining the smeared delta function while the parameter $A$ gets two different values, $A_b$ in case of  baryons comprising at least a  beauty and $A_c$ otherwise. In the one-gluon-exchange approximation, the parameter $A$ is proportional to the strong coupling constant $\alpha_s$, therefore an argument supporting the two values $A_c$ and $A_b$ is represented by the scales, $O(m_c)$ and $O(m_b)$, to which $\alpha_s$ must be computed in the two cases.

 A cut-off at small distance is introduced to cure the singularity of the wave function; it consists in fixing the potential (\ref{potenziale}) at the value ${V}(r_M)$ for $r\leq r_M$, with $r_M=\frac{4\Lambda\pi}{3 M}$ \cite{Cea:1986bj}, $M$ being the mass of the diquark and  $\Lambda$ a parameter; $\Lambda=1$ in case of $m_1=m_2$, as discussed in  \cite{Colangelo:1990rv}.

\begin{figure}[h]
\begin{center}
\includegraphics[width=8cm]{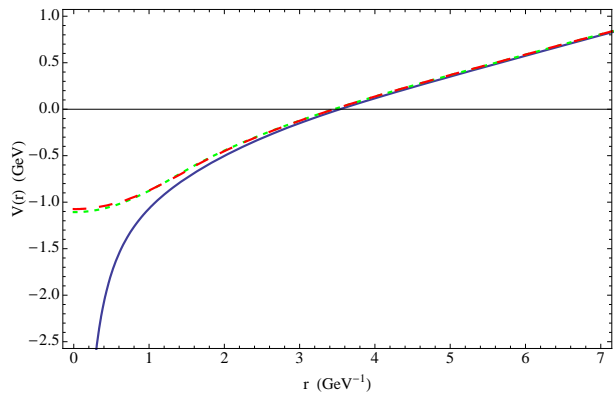}
\caption{Quark-diquark potential $\tilde V(r)$, for the  $\{cc\}$ diquark in the 1$S$ state (dashed line) and for the $\{cc\}$ diquark in the 2$S$ state (dotted line),  and quark-antiquark potential $V(r)$ (\ref{potenziale}) (continuous line).}
\label{figpot}
\end{center}
\end{figure}

Once the diquark masses have been obtained, one can use the Salpeter equation to study the interaction between a diquark and a quark, obtaining the baryon masses. As already stated, the energy of a quark-diquark pair is assumed to be the same as the one of a quark-antiquark pair: this suggests to adopt again the potential (\ref{potenziale}). However,  diquarks are extended objects: therefore, to keep this into account we construct  the potential using a convolution with the diquark wave function: 
\begin{equation}
 \tilde V(R)=\frac1{N}\int d{\bf r} \; |\psi_{d}({\bf r})|^2
V( |{\bf R}+{\bf r}|)
\label{barpot}
\end{equation}
where $\psi_{d}$ is the wave function of the diquark and $N$ is a normalization factor. 
The integral (\ref{barpot}) runs from $r=0$ to a radius $\displaystyle r_{max}$ which ensures that 
the  diquark is on average inside the baryon's bag.
The obtained potential $\tilde V(r)$ is in Fig. \ref{figpot}, together with the quark-antiquark potential (\ref{potenziale}) (continuous  line): the dashed line represents the potential obtained through the 1$S$ wave function of the diquark $\{cc\}$, while the dotted line represents the potential obtained through the 2$S$ wave function of the diquark $\{cc\}$ ($\{cc\}$ indicates a spin 1 diquark with two charm quarks). The figure shows that  a  similar potential is obtained for the interaction between a quark and a diquark, when the diquark is in the 1$S$ or 2$S$ state.

The Salpeter equation for a baryon can be finally written in the following way:
\begin{equation}\label{barsalpetereq}
\left(\sqrt{m_q^2-\nabla^2}+\sqrt{m_{d}^2-\nabla^2}+\tilde{V}(r)\right)
\psi({\bf r})\,=\,M\, \psi({\bf r})\, ,
\end{equation}
where $m_q$ is the mass of the constituent quark,  $m_d$ is the mass of the constituent  diquark and $M$ and $\psi({\bf r})$ are the mass and the wave function of the baryon, respectively. Again, we only consider the $\ell$=0 case  for the system quark-diquark.

The Salpeter equations (\ref{salpetereq}) and (\ref{barsalpetereq}) can be solved through the Multhopp method \cite{Colangelo:1990rv}.
The parameters of the model, as in ref. \cite{Giannuzzi}, are:  $c=$0.4 GeV$^2$, $g$=2.50, $V_0$=-0.47 GeV, $A_c$=14.56, $A_b$=6.49,  $\sigma$=0.47 GeV,  $\Lambda=0.5$ in the potential, and the constituent quark masses $m_q$=0.34 GeV ($q=u,d$), $m_s$=0.48 GeV, $m_c$=1.59 GeV and $m_b$=5.02 GeV: these values have been obtained  by a best fit of the meson masses computed in this model to their experimental values  \cite{pdg}.

 The values obtained for diquark masses are shown in Table \ref{tabdiquark} for the 1$S$ and 2$S$ states.  A diquark with spin 1 is denoted by $\{QQ\}$, while a diquark with spin 0 is denoted by $[QQ]$. Notice that a diquark with two identical quarks in $\bar 3$ can only have spin 1, as far as the $\ell=0$ case is considered, in order to make the wave function of the two quarks antisymmetric  \cite{Jaffe:2004ph}.  

\begin{table}[ht!]
\caption{Diquark masses in GeV.  $\{QQ\}_{nS}$ (resp. $[QQ]_{nS}$) means a spin 1 (resp. spin 0)
diquark $QQ$ in $S$ wave with radial number $n$.} 
\begin{center}
\begin{tabular}{|c|c|c|}\hline
Diquark& State& Mass \\
\hline $\{cc\}_{nS}$&1$S$ & 3.238 \\
& 2$S$ & 3.589\\
\hline $[bc]_{nS}$& 1$S$ &6.558\\
& 2$S$ & 6.882 \\
\hline $\{bc\}_{nS}$&1$S$ & 6.562  \\
& 2$S$ & 6.883 \\
\hline $\{bb\}_{nS}$&1$S$ &9.871 \\
& 2$S$ & 10.165\\
\hline
\end{tabular}
\end{center}
\label{tabdiquark}
\end{table}

\begin{figure}[h]
\subfigure{
\includegraphics[scale=0.51]{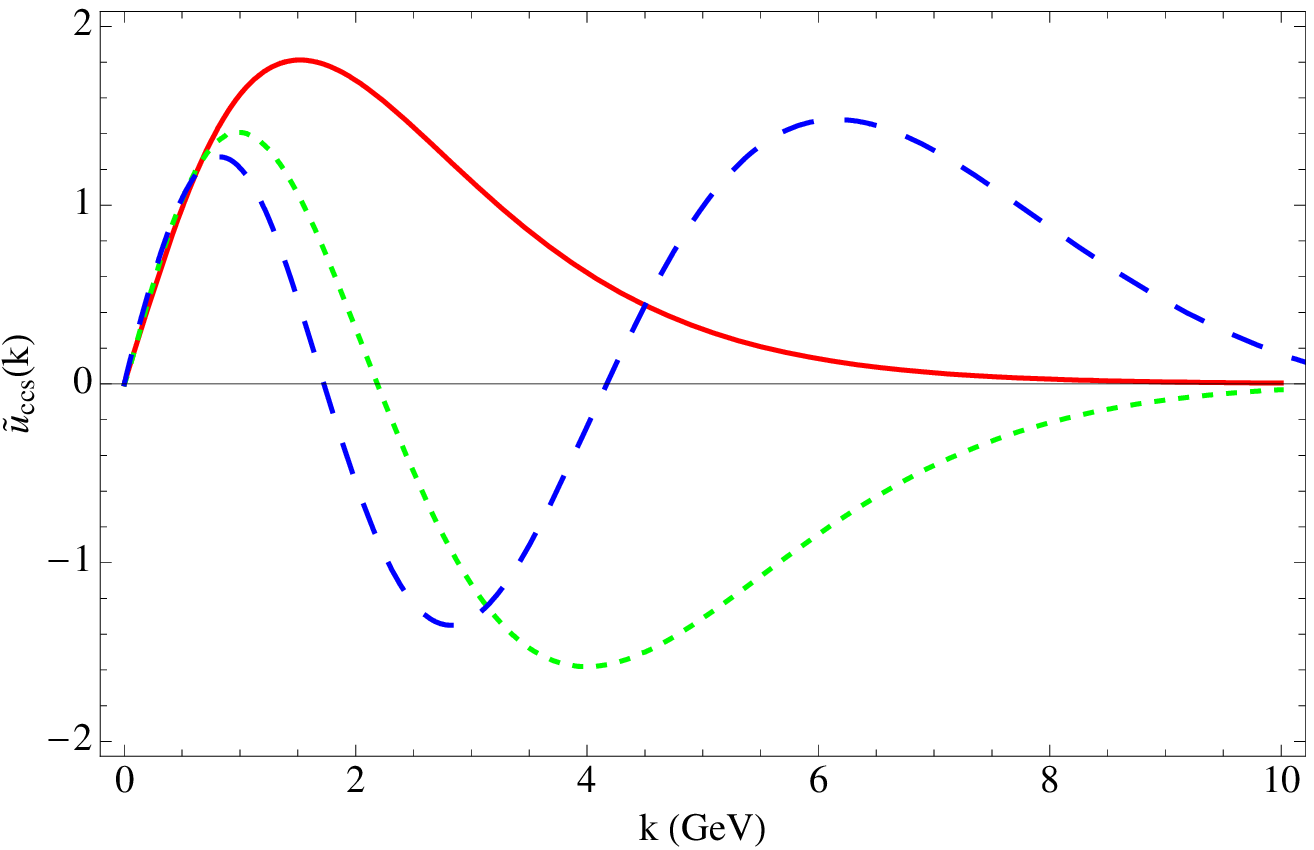}
}
\hspace*{1cm}
\subfigure{
\includegraphics[scale=0.52]{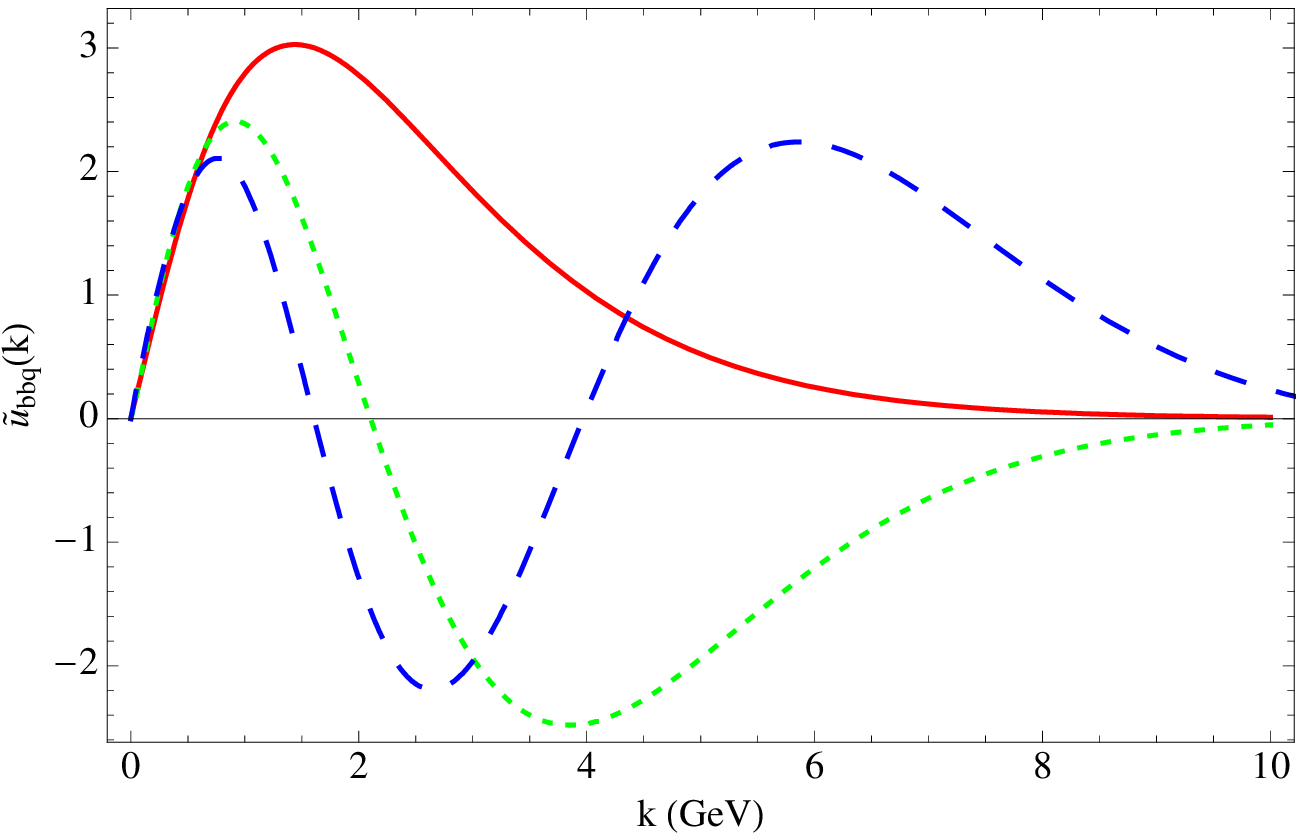}
}
\caption{Wave functions of the first three radial excitations of $\Omega_{cc}$ (left) and $\Xi_{bb}$ (right). The continuous line represents the 1$S$ wave function, the dotted line the 2$S$ wave function and the dashed line the 3$S$ wave function.
The wave functions are dimensionless: they are normalized as $\int dk \, |\tilde{u}(k)|^2=2M$, being $k$ the modulus of the relative 3-momentum of the quark-diquark pair. }
\label{wfunc}
\end{figure}

The masses of doubly heavy baryons are shown in Table \ref{dhbaryonscc} for baryons with a $\{cc\}_{1S}$ diquark, in Table \ref{dhbaryonsbb} for baryons with a $\{bb\}_{1S}$ diquark, and in Table \ref{dhbaryonscb} for baryons with a $[bc]_{1S}$ or $\{bc\}_{1S}$ diquark. The results are compared with recent models: ref. \cite{Valcarce:2008dr,Roberts:2007ni,Albertus:2006ya} describe baryons by a non-relativistic quark model  based on a three-body problem; in ref. \cite{Ebert:2002ig,Kiselev:2001fw} potential models  based on the quark-diquark hypothesis are investigated, the first one relativistic and the second one non-relativistic; in ref.  \cite{Zhang:2008rt} doubly heavy baryon masses are computed in the framework of QCD sum rules; ref. \cite{Mathur:2002ce,Flynn:2003vz} deal with quenched lattice QCD, and finally ref. \cite{Bernotas:2008fv} is based on the bag model. In Fig. \ref{wfunc} the wave functions of the first three radial excitations of $\Omega_{cc}$ and $\Xi_{bb}$ are shown. Since $\ell=0$, all the states have positive parity.

\begin{table}[h]
\caption{Masses (GeV) of baryons composed by a diquark in the lowest mass configuration $\{cc\}_{1S}$ and a light quark ($q$ or $s$).  In the case of ref. \cite{Mathur:2002ce}, the first and the second results are obtained using $\beta=2.1$ and  $\beta=2.3$, respectively. }
\begin{center}
\begin{tabular}{|c|c|c|c|c|c|c|c|c|c|c|c|c|c|}
\hline
Particle & State & $J^P$& Quark-diquark content & This paper &\cite{Valcarce:2008dr}&\cite{Roberts:2007ni} &  \cite{Albertus:2006ya}&\cite{Ebert:2002ig}&\cite{Kiselev:2001fw} & \cite{Zhang:2008rt} &\cite{Mathur:2002ce} &\cite{Flynn:2003vz} & \cite{Bernotas:2008fv}\\
\hline
$\Xi_{cc}$ & 1$S$&  $\frac{1}{2}^+$&$q\{cc\}_{1S}$ & 3.547 & 3.579 & 3.676 & 3.612 & 3.620 & 3.48 & 4.26 &3.562 (3.588) & 3.549 & 3.557  \\
& 2$S$&&& 4.183 & 3.876 &&  &&&&&& \\
& 3$S$&&& 4.640 &&&&&&&&&\\
\hline
$\Xi_{cc}^*$  & 1$S$& $\frac{3}{2}^+$&  $q\{cc\}_{1S} $ &3.719 & 3.656  &3.753 & 3.706 & 3.727& 3.61 & 3.90& 3.625 (3.658) & 3.641 & 3.661\\
& 2$S$&&& 4.282 &4.025  &&&&&&&&\\
&3$S$&& &4.719 &&&&&&&&&\\
\hline
$\Omega_{cc}$  & 1$S$ & $\frac{1}{2}^+$ &$s \{cc\}_{1S}$  & 3.648 & 3.697  & 3.815 & 3.702& 3.778& 3.59 & 4.25 & 3.681 (3.698) & 3.663 &3.710 \\
& 2$S$&&& 4.268 & 4.112 &&&&&&&&\\
 &3$S$&&& 4.714 &&&&&&&&&\\
\hline
$\Omega_{cc}^*$  &  1$S$ & $\frac{3}{2}^+$ & $s\{cc\}_{1S}$ & 3.770 & 3.769   & 3.876 & 3.783& 3.872& 3.69 & 3.81 & 3.737 (3.761) & 3.734 & 3.800\\
& 2$S$&&& 4.334 & & &&&&&&&\\  
& 3$S$&&& 4.766 &&&&&&&&&\\
\hline
\end{tabular}
\end{center}
\label{dhbaryonscc}
\end{table}%

\begin{table}[h]
\caption{Masses (GeV) of baryons composed by a diquark $\{bb\}_{1S}$ and a light quark ($q$ or $s$).}
\begin{center}
\begin{tabular}{|c|c|c|c|c|c|c|c|c|c|c|c|c|c|}
\hline
Particle & State & $J^P$& quark-diquark content & This paper &\cite{Valcarce:2008dr} &\cite{Roberts:2007ni} &  \cite{Albertus:2006ya}&\cite{Ebert:2002ig}&\cite{Kiselev:2001fw} & \cite{Zhang:2008rt}&\cite{Lewis:2008fu}& \cite{Bernotas:2008fv}\\
\hline
$\Xi_{bb}$ & 1$S$& $\frac{1}{2}^+$&$q \{bb\}_{1S}$  &  10.185 & 10.189   & 10.340 & 10.197 & 10.202& 10.09 & 9.78 & 10.127 & 10.062 \\
& 2$S$&&& 10.751 & 10.586& & &&&&&\\
& 3$S$&&& 11.170 &&&&&&&&\\
\hline
$\Xi_{bb}^*$ & 1$S$ & $\frac{3}{2}^+$ &$q \{bb\}_{1S}$ & 10.216 & 10.218 & 10.367 & 10.236& 10.237& 10.13 & 10.35 & 10.151 & 10.101 \\
& 2$S$&&& 10.770 & 10.501 & &&&&&&\\
& 3$S$&&& 11.184 &&&&&&&&\\
\hline
$\Omega_{bb}$ & 1$S$ & $\frac{1}{2}^+$& $s\{bb\}_{1S}$ &10.271 & 10.293    &10.454 & 10.260& 10.359 & 10.18 & 9.85 & 10.225 & 10.208\\
& 2$S$&&& 10.830 & 10.604 & &&&&&&\\
& 3$S$&&& 11.240 &&&&&&&&\\
\hline
$\Omega_{bb}^*$ & 1$S$ & $\frac{3}{2}^+$ & $s\{bb\}_{1S}$ &  10.289 & 10.321 &10.486 & 10.297 & 10.389 & 10.20 & 10.28 & 10.246 & 10.244\\
& 2$S$&&& 10.839 & 10.622   &&&&&&&\\
& 3$S$&&& 11.247 &&&&&&&&\\
\hline
\end{tabular}
\end{center}
\label{dhbaryonsbb}
\end{table}%

\begin{table}[h]
\caption{Masses (GeV) of baryons composed by a diquark $bc$ in the lowest mass configuration and a light quark ($q$ or $s$).}
\begin{center}
\begin{tabular}{|c|c|c|c|c|c|c|c|c|c|c|}
\hline
Particle & State & $J^P$&Quark-diquark content & This paper   &\cite{Roberts:2007ni} &  \cite{Albertus:2006ya}&\cite{Ebert:2002ig}&\cite{Kiselev:2001fw} & \cite{Zhang:2008rt} & \cite{Bernotas:2008fv} \\
\hline
$\Xi_{bc}$ & 1$S$ & $\frac{1}{2}^+$ &$q\{bc\}_{1S}$ & 6.904  & 7.011 & 6.919 & 6.933 & 6.82 & 6.75 &6.846  \\
& 2$S$&&& 7.478  &&&&&&\\
& 3$S$&&& 7.904 &&&&&&\\
\hline
$\Xi_{bc}'$ & 1$S$ & $\frac{1}{2}^+$ &$q[bc]_{1S}$& 6.920   & 7.047 & 6.948 & 6.963 & 6.85 & 6.95& 6.891 \\
& 2$S$&&& 7.485  &&&&&&\\
& 3$S$&&& 7.908 &&&&&&\\
\hline
$\Xi_{bc}^*$ & 1$S$& $\frac{3}{2}^+$&$q\{bc\}_{1S}$& 6.936   & 7.074 & 6.986 & 6.980 & 6.90 & 8.00 & 6.919 \\
& 2$S$&&& 7.495  &&&&&&\\
& 3$S$&&& 7.917 &&&&&&\\
\hline
$\Omega_{bc}$ & 1$S$& $\frac{1}{2}^+$ & $s\{bc\}_{1S}$ &  6.994   & 7.136 & 6.986 & 7.088 & 6.91& 7.02  & 6.999\\
& 2$S$&&& 7.559  &&&&&&\\
& 3$S$&&& 7.976 &&&&&&\\
\hline
$\Omega_{bc}'$ & 1$S$ & $\frac{1}{2}^+$ & $s[bc]_{1S}$ & 7.005   & 7.165 & 7.009& 7.116 & 6.93 & 7.02 & 7.036 \\
& 2$S$&&& 7.563  &&&&&&\\
& 3$S$&&& 7.977 &&&&&&\\
\hline
$\Omega_{bc}^* $& 1$S$ & $\frac{3}{2}^+$ &$s\{bc\}_{1S}$ &  7.017   & 7.187 & 7.046& 7.130 & 6.99 & 7.54 & 7.063 \\
& 2$S$&&& 7.571  &&&&&&\\
& 3$S$&&& 7.985 &&&&&&\\
\hline
\end{tabular}
\end{center}
\label{dhbaryonscb}
\end{table}%

A few remarks are in order. First, the value found in this paper for the mass of $\Xi_{cc}$ is in agreement with the experimental value found by SELEX Collaboration (\ref{csiccselex}), taking into account the uncertainties in the quark masses and those related to our description of the baryon. In fact, the difference between the experimental and the theoretical value of the mass is of the same order than the differences found for meson masses in \cite{Giannuzzi}.\\
The only remarkable difference between our results and the others shown in  the tables concerns the radial excitations, since the masses evaluated within this paper are higher than the ones found in ref. \cite{Valcarce:2008dr}, which could be due to the different value of the string tension: however, the parameters in our approach are fixed by a best fit of meson masses, including radial resonances of $J/\psi$ and $\Upsilon$. 

In  \cite{Roberts:2007ni} it was argued that the first excited state of a baryon comprising a quark and a heavy diquark is the one with the diquark  in an excited state, namely the 2$S$ state: this  level could be lower than the one corresponding to the 2$S$ radial excitation of the whole baryon. The masses of baryons  with the diquark in the $2S$ state computed in our approach  are shown in Table \ref{diqexcbaryons}, together with the results of other models.  The masses we have obtained  are comparable with the values found within the other models. 
Concerning baryons in Table \ref{dhbaryonscb}, these excited levels are not reported because  the excited states of diquarks $\{bc\}$ and $[bc]$ are not stable due to the emission of soft gluons \cite{Kiselev:2001fw}.

\begin{table}[h]
\caption{Masses (GeV) of the excited baryons in which the diquark is in the 2$S$ state.}
\begin{center}
\begin{tabular}{|c|c|c|c|c|c|c|}
\hline
Baryon & $J^P$& Quark-diquark content& This paper & \cite{Roberts:2007ni} & \cite{Ebert:2002ig}&  \cite{Kiselev:2001fw,Gershtein:2000nx} \\
\hline
  $\Xi_{cc}$ & $\frac{1}{2}^+$& $q\{cc\}_{2S}$& 3.893 & 4.029 &  3.910&  3.812\\
 \hline
 $\Xi_{cc}^*$& $\frac{3}{2}^+$&$q\{cc\}_{2S}$&4.021& 4.042 & 4.027& 3.944\\
 \hline
 $\Omega_{cc}$& $\frac{1}{2}^+$&$s\{cc\}_{2S}$& 3.992&  4.180 & 4.075& \\
 \hline
 $\Omega_{cc}^*$& $\frac{3}{2}^+$&s$\{cc\}_{2S}$& 4.105&  4.188  & 4.174& \\
 \hline
 $\Xi_{bb}$& $\frac{1}{2}^+$& $q\{bb\}_{2S}$&10.453&  10.576 &  10.441& 10.373\\
 \hline
 $\Xi_{bb}^*$& $\frac{3}{2}^+$& $q\{bb\}_{2S}$& 10.478& 10.578 &  10.482& 10.413\\
 \hline
 $\Omega_{bb}$& $\frac{1}{2}^+$&$s\{bb\}_{2S}$& 10.538& 10.693 & 10.610& \\
 \hline
 $\Omega_{bb}^*$ & $\frac{3}{2}^+$& $s\{bb\}_{2S}$&  10.556 & 10.721 &  10.645 & \\
  \hline
\end{tabular}
\end{center}
\label{diqexcbaryons}
\end{table}%

It is interesting to analyze the results using the language of HQET. Analogously to the $1/m_Q$ expansion of the mass of a baryon comprising a single heavy quark  \cite{Jenkins:1996de}, one can attempt to write an expansion with respect to  the inverse of the heavy diquark mass for a baryon made up of  a heavy diquark and a light quark:
\begin{equation}\label{hqetbaryon}
M_{\{QQ\}q}=m_{\{QQ\}}+\bar \Lambda+\frac{\lambda_1}{2m_{\{QQ\}}}+A_Q d_H \frac{\lambda_2}{2 m_{\{QQ\}}}
\end{equation}
where $m_{\{QQ\}}$ is the mass of the diquark and $d_H$ is $d_H={\bf S}_{\{QQ\}}\cdot {\bf S}_q$. The mass splitting between $J^P=3/2^+$ and $J^P=1/2^+$ baryons turns out to be, for example in case of $\Xi_{QQ}$:
\begin{equation}\label{hqetbsplit}
\Xi_{QQ}^*-\Xi_{QQ}=A_Q \frac{3 \lambda_2}{4 m_{\{QQ\}}} \,.
\end{equation}
From Eq. (\ref{hqetbsplit}), the ratio between the mass splitting of $\Xi_{bb}$ and $\Xi_{cc}$ and between the difference of the mass squared:
\begin{equation}
\frac{\Xi_{bb}^*-\Xi_{bb}}{\Xi_{cc}^*-\Xi_{cc}}=\frac{A_b m_{\{cc\}}}{A_c m_{\{bb\}}}  \qquad , \qquad \frac{ \Xi_{bb}^{*2}-\Xi_{bb}^2}{\Xi_{cc}^{*2}-\Xi_{cc}^2}=\frac{A_b}{A_c}  \,,
\end{equation}
relations well verified, both for $\Xi_{QQ}$ and for $\Omega_{QQ}$ baryons, as one can appreciate considering the results  in Table \ref{dhbaryonscc} and \ref{dhbaryonsbb}. Moreover, a mass splitting hierarchy is obtained:
$$(\Xi_{cc}^*-\Xi_{cc})>(\Omega_{cc}^*-\Omega_{cc})>(\Xi_{bb}^*-\Xi_{bb})>(\Omega_{bb}^*-\Omega_{bb}) \,.$$

As a final result, we collect in Table \ref{thbaryons} the masses of baryons with three heavy quarks. However, we point out that such last  predictions, obtained substituting the third quark with a heavy one, have to be considered with caution, since, although the presence of  heavy interacting particles is a preferable condition for the application of the static potential (\ref{potenziale}), the hypothesis of a quark-diquark configuration becomes less reliable when the average distances between each pair of quarks are comparable.

\begin{table}[h]
\caption{Masses (GeV) of baryons made up of a diquark $\{cc\}$ or $\{bb\}$ and a heavy quark ($c$ or $b$).}
\begin{center}
\begin{tabular}{|c|c|c|c|c|c|c|c|c|c|c|}
\hline
Particle & State& $J^P$&Quark-diquark content & This paper & \cite{Zhang:2009re} & \cite{Bernotas:2008fv} & \cite{Roberts:2007ni} &\cite{Kiselev:2001fw} \\
\hline
$\Omega_{ccb} $& 1$S$&$\frac{1}{2}^+$ & $b\{cc\}_{1S}$ & 7.832& 7.41 & 7.984  &8.245   &   \\
& 2$S$&&& 8.350  &&& 8.537&\\
& 3$S$&&& 8.704 &&&&\\
\hline
$\Omega_{ccb}^* $& 1$S$&$\frac{3}{2}^+$&$b\{cc\}_{1S}$ & 7.839& 7.45 & 8.005 & 8.265   &  \\
& 2$S$&&& 8.353  &&&8.553&\\
& 3$S$&&& 8.706 &&&&\\
\hline
$\Omega_{bbc} $& 1$S$&$\frac{1}{2}^+$&$c\{bb\}_{1S}$& 11.108& 10.30 & 11.139 & 11.535  & 11.12    \\
& 2$S$&&& 11.639  &&& 11.787&\\
& 3$S$&&& 12.010 &&&&\\
\hline
$\Omega_{bbc}^* $& 1$S$&$\frac{3}{2}^+$&$c\{bb\}_{1S}$& 11.115 & 10.54 & 11.163 & 11.554& 11.18     \\
& 2$S$&&& 11.642&&  &11.798&\\
& 3$S$&&& 12.012 &&&&\\
\hline
\end{tabular}
\end{center}
\label{thbaryons}
\end{table}%

Baryons with two and three heavy quarks complete the set of states predicted by the quark model for ordinary hadrons. Only one state, the lightest one $\Xi_{cc}$, has been observed so far, but  the existence of the other baryons  could be proved by forthcoming experiments. Models can be constructed to  predict their masses: the model described in this paper uses the scheme of a quark-diquark configuration for doubly heavy baryons and is completely defined by fitting the meson spectrum. The obtained values are in agreement with the only known experimental result.

\vspace{1cm}

{\bf Acknowledgements}

I would like to thank   P.~Colangelo, F.~De Fazio and S.~Nicotri  for collaboration and  precious suggestions and discussions. I also thank M.V. Carlucci, M. Pellicoro and S. Stramaglia  for collaboration on developing the numerical method used here. This work was supported in part by the EU Contract No. MRTN-CT-2006-035482, "FLAVIAnet".


\begin{thebibliography}{99}
\bibitem{Fleck:1988vm}
  S.~Fleck, B.~Silvestre-Brac and J.~M.~Richard,
  Phys.\ Rev.\  D {\bf 38}, 1519 (1988).


\bibitem{Mattson:2002vu}
  M.~Mattson {\it et al.}  [SELEX Collaboration],
  Phys.\ Rev.\ Lett.\  {\bf 89}, 112001 (2002)
  [arXiv:hep-ex/0208014].

\bibitem{Ocherashvili:2004hi}
  A.~Ocherashvili {\it et al.}  [SELEX Collaboration],
  Phys.\ Lett.\  B {\bf 628}, 18 (2005)
  [arXiv:hep-ex/0406033].
  
  \bibitem{Kiselev:2001fw}
  V.~V.~Kiselev and A.~K.~Likhoded,
  Phys.\ Usp.\  {\bf 45}, 455 (2002)
  [Usp.\ Fiz.\ Nauk {\bf 172}, 497 (2002)]
  [arXiv:hep-ph/0103169].
  
\bibitem{Liu:2007twb}
  X.~Liu, H.~W.~Ke, Q.~P.~Qiao, Z.~T.~Wei and X.~Q.~Li,
  Phys.\ Rev.\  D {\bf 77}, 035014 (2008)
  [arXiv:0710.2600 [hep-ph]].
  
\bibitem{Ida:1966ev}
  M.~Ida and R.~Kobayashi,
  Prog.\ Theor.\ Phys.\  {\bf 36}, 846 (1966);
D.~B.~Lichtenberg and L.~J.~Tassie,
  Phys.\ Rev.\  {\bf 155}, 1601 (1967).
  
\bibitem{Klempt:2009pi}
  E.~Klempt and J.~M.~Richard,
  arXiv:0901.2055 [hep-ph].

\bibitem{Carlucci:2007um}
 M.~V.~Carlucci, F.~Giannuzzi, G.~Nardulli, M.~Pellicoro and S.~Stramaglia,
  Eur.\ Phys.\ J.\  C {\bf 57}, 569 (2008)
  [arXiv:0711.2014 [hep-ph]].
  
  \bibitem{Giannuzzi}
  F.~Giannuzzi,
  Phys.\ Rev.\  D {\bf 78}, 117501 (2008)
  [arXiv:0810.2736 [hep-ph]].

\bibitem{Andreev:2006ct}
  O.~Andreev and V.~I.~Zakharov,
  Phys.\ Rev.\  D {\bf 74}, 025023 (2006)
  [arXiv:hep-ph/0604204].
  
\bibitem{Cea:1986bj}
  P.~Cea and G.~Nardulli,
  Phys.\ Rev.\  D {\bf 34}, 1863 (1986).

\bibitem{Colangelo:1990rv}
  P.~Colangelo, G.~Nardulli and M.~Pietroni,
  Phys.\ Rev.\  D {\bf 43}, 3002 (1991).

\bibitem{pdg}
C.~Amsler {\it et al.} [Particle Data Group], Phys. Lett. B {\bf 667}, 1 (2008).

\bibitem{Jaffe:2004ph}
  R.~L.~Jaffe,
  Phys.\ Rept.\  {\bf 409}, 1 (2005)
  [Nucl.\ Phys.\ Proc.\ Suppl.\  {\bf 142}, 343 (2005)]
  [arXiv:hep-ph/0409065].

\bibitem{Valcarce:2008dr}
  A.~Valcarce, H.~Garcilazo and J.~Vijande,
  Eur.\ Phys.\ J.\  A {\bf 37}, 217 (2008)
  [arXiv:0807.2973 [hep-ph]].


\bibitem{Roberts:2007ni}
  W.~Roberts and M.~Pervin,
  Int.\ J.\ Mod.\ Phys.\  A {\bf 23}, 2817 (2008)
  [arXiv:0711.2492 [nucl-th]].


\bibitem{Albertus:2006ya}
  C.~Albertus, E.~Hernandez, J.~Nieves and J.~M.~Verde-Velasco,
  Eur.\ Phys.\ J.\  A {\bf 32}, 183 (2007)
  [Erratum-ibid.\  A {\bf 36}, 119 (2008)]
  [arXiv:hep-ph/0610030].

\bibitem{Ebert:2002ig}
  D.~Ebert, R.~N.~Faustov, V.~O.~Galkin and A.~P.~Martynenko,
  Phys.\ Rev.\  D {\bf 66}, 014008 (2002)
  [arXiv:hep-ph/0201217].

\bibitem{Zhang:2008rt}
  J.~R.~Zhang and M.~Q.~Huang,
  Phys.\ Rev.\  D {\bf 78}, 094007 (2008)
  [arXiv:0810.5396 [hep-ph]].

\bibitem{Mathur:2002ce}
  N.~Mathur, R.~Lewis and R.~M.~Woloshyn,
  Phys.\ Rev.\  D {\bf 66}, 014502 (2002)
  [arXiv:hep-ph/0203253].

\bibitem{Flynn:2003vz}
  J.~M.~Flynn, F.~Mescia and A.~S.~B.~Tariq  [UKQCD Collaboration],
  JHEP {\bf 0307}, 066 (2003)
  [arXiv:hep-lat/0307025].

\bibitem{Bernotas:2008fv}
  A.~Bernotas and V.~Simonis,
  arXiv:0801.3570 [hep-ph].

\bibitem{Lewis:2008fu}
  R.~Lewis and R.~M.~Woloshyn,
  Phys.\ Rev.\  D {\bf 79}, 014502 (2009)
  [arXiv:0806.4783 [hep-lat]].

\bibitem{Jenkins:1996de}
  E.~E.~Jenkins,
  Phys.\ Rev.\  D {\bf 54}, 4515 (1996)
  [arXiv:hep-ph/9603449].

\bibitem{Gershtein:2000nx}
  S.~S.~Gershtein, V.~V.~Kiselev, A.~K.~Likhoded and A.~I.~Onishchenko,
  Phys.\ Rev.\  D {\bf 62}, 054021 (2000).

\bibitem{Zhang:2009re}
  J.~R.~Zhang and M.~Q.~Huang,
  arXiv:0902.3297 [hep-ph].














\end{thebibliography}
\end{document}